\documentclass{iaus}
\usepackage{graphicx}

\title[Jets in pre-planetary nebulae] 
{Jet power in pre-planetary nebulae: \\
observations vs.\ theory}

\author[P. J. Huggins]   
{P. J. Huggins}

\affiliation{Physics Department, New York University, \\
4 Washington Place, New York NY 10003, USA\\ 
email: {\tt patrick.huggins@nyu.edu} \\
}

\pubyear{2011}
\volume{283}  
\pagerange{119--126}
\jname{Planetary Nebulae: an Eye to the Future}
\editors{A. Manchado, L. Stanghellini \& D. Schoenberner, eds.}
\begin{document}

\maketitle

\begin{abstract}
High velocity jets are among the most prominent features of a wide
class of planetary nebulae, but their origins are not understood.
Several different types of physical model have been suggested to power
the jets, but there is no consensus or preferred scenario.  We
compare current theoretical ideas on jet formation with observations,
using the best studied pre-planetary nebulae in millimeter CO, where
the dynamical properties are best defined. In addition to the mass,
velocity, momentum, and energy of the jets, the mass and energetics of
the equatorial mass-loss that typically accompanies jet formation
prove to be important diagnostics. Our integrated approach provides
estimates for some key physical quantities - such as the binding
energy of the envelope when the jets are launched - and allows testing
of model features using correlations between parameters. Even with a
relatively small sample of well-observed objects, we find that some
specific scenarios for powering jets can be ruled out or rendered
implausible, and others are promising at a quantitative level.

\end{abstract}

\firstsection 
\section{Introduction}

Jets are prominent and common features of planetary nebulae
(PNe). They are known to develop in the rapid transition from the
asymptotic giant branch (AGB) to the pre-PN phase, but their formation
is not well understood. This paper outlines an approach to constrain
the origins of the jets by confronting basic jet theory with
observations. This is a challenging objective because the regions
where the jets form are star-sized, and cannot be resolved by current
observations. Thus the formation mechanisms have to be identified
using properties that can be observed on much larger size scales.

From the earliest CO surveys of evolved stars, it was recognized that
the mass-loss of pre-PNe is different from that of AGB stars
(\cite[Knapp et al.\ 1982]{kna82}). Broad velocity wings were later
discovered in the CO spectra which were identified with optical jets, and
over ten years ago, \cite[Bujarrabal et al.\ (2001)]{buj01} showed
that these could not be powered by the radiation field of the central star.

Since that time, an important development has been the ability to map
the CO emission of individual objects in more detail (e.g., \cite[Cox
et al.\ 2000]{cox00}; \cite[Alcolea et al.\ 2007]{alc07}).  Together
with HST imaging (e.g., \cite[Sahai et al.\ 2007]{sah07}) a standard
picture has emerged of a typical pre-PN. It consists of an extended
remnant AGB envelope, an inner region of enhanced mass-loss which
appears as a torus, and jets which consist mainly of entrained
material.

The jets and tori together constitute the last major mass-loss in the
formation of a PN. Hence their timing, geometry, and dynamics all
offer potentially important clues to the ejection process. The timing
has recently been studied by \cite[Huggins (2007)]{hug07}, who finds
that jets and tori are quasi-simultaneous, with evidence for a
preferred sequence. This paper deals with their dynamical properties.

\section{Observations} 
The approach adopted here is to assemble a sample of the best
available CO observations of individual pre-PNe with strong jets to
investigate the systematics of the ensemble and to compare them with
theoretical expectations.

The sample consists of 12 bona fide pre-PNe for which good estimates
can be made of the mass, velocity, momentum, and energy of the jets
and tori. The spectral types of the central stars range from M to
Be. The objects are: IRAS 04395$+$3601, 09371$+$1212, 11385$-$5517,
17423$-$1755, 17436$+$5003 19343$+$2926, 19475$+$3119, 19500$-$1709,
22036 $+$5306, 23304$+$6147, 23541$+$7031, and AFGL 2688. The
observations are the combined work of several groups -- the main
sources include: 
\cite[Alcolea et al.\ (2007)]{alc07}, 
\cite[Bujarrabal et al.\ (2001)]{buj01}, 
\cite[Castro-Carrizo et al.\ (2002)]{cas02},
\cite[Castro-Carrizo et al.\ (2005)]{cas05}, 
\cite[Olofsson \& Nyman (1999)]{olo99}, 
\cite[Sahai et al.\ (2006)]{sah06},  
\cite[S{\'a}nchez Contreras et al.\ (2004)]{san04}, 
and \cite[S{\'a}nchez Contreras et al.\ (2006)]{san06}.

\section{Dynamical properties}
We discuss in turn the dynamical properties of the jets and tori that
can be derived from the observations. Quantities that involve the mass
vary as distance squared. The distance in most cases is not very well
known, but the main conclusions do not depend on this uncertainty.

{\underline{\it Masses}}.  We first check the observed masses of the
jets and tori to assess their contribution to the mass budget. Their
combined mass, $m_{TOT}$, which is the mass of the AGB envelope at the
final ejection phase, ranges from 0.05 to 1.0~M$_\odot$. This is
roughly centered on the typical mass of a mature, ionized PN. Hence
the jets and tori make a significant contribution to the masses of
PNe. The mass fraction in the jets ranges from 0.08 to 1, consistent
with the range in solid angle of typical jet structures seen in
optical images.

{\underline{\it Velocities}}.  The mean velocities of the tori range
from 7 to 16~km s$^{-1}$, and the mean velocities of the jets range
from 22 to 150~km s$^{-1}$.  The maximum velocities of the jets are
approximately a factor of three higher.

It is found that the jet velocities decrease with increasing envelope
mass ($m_{TOT}$), in accord with the idea that the outflows are
powered by light jets that are mass-loaded by the entrainment
of material. It is also found -- unexpectedly -- that the mean
velocities of the tori correlate with the mean velocities of the jets,
as shown in Fig.~1 (left). This implies that jets and tori are
dynamically coupled. In fact, the relation extrapolates to small or
zero torus velocity for zero jet velocity, which suggests that the
tori may actually be driven from the vicinity of the central star by
the jets.

{\underline{\it Kinetic energies}}.  The observed kinetic energies of
the outflows are plotted vs.\ $m_{TOT}$ in Fig.~1 (right). The tori and
jets are shown separately: it can be seen that the jets dominate the
energetics.

Fig.~1 includes curves for two reference energies. The first, labeled
$E_\mathrm{sec}$, is the maximum energy that could be extracted by
accretion of the whole AGB envelope onto a main sequence secondary
star. This is given by the usual expression for accretion energy
$GM_\mathrm{sec}m_\mathrm{acc}/2r_\mathrm{sec}$, with the accretion
mass $m_\mathrm{acc}$ replaced with $m_{TOT}$. Note that the ratio
$M_\mathrm{sec}/r_\mathrm{sec}$ for stars on the lower main sequence
is approximately constant, so that $E_\mathrm{sec}$ is essentially
independent of the mass of the companion and depends only on the
accreted mass.

The second reference energy, $E_\mathrm{bin}$, is the binding energy
of an AGB star with envelope mass $m_{TOT}$, assuming a core mass of
0.6~M$_\odot$, an AGB radius of 350~R$_\odot$, and a binding
coefficient $\lambda = 0.25$. If the jets are somehow powered by the
infall of a companion, a standard assumption of common envelope
calculations is that the infall terminates when enough mechanical
energy is generated to unbind the envelope. In this case we expect the
observed kinetic energy to be $< E_\mathrm{bin}$.

From Fig.~1 it can be seen that the energies of the jets and tori are
significantly less than $E_\mathrm{sec}$. The energies of the tori are
also significantly less than $E_\mathrm{bin}$. This is consistent with several
possibilities for torus formation (including the standard picture of
common envelope ejection) in which the final kinetic energy is small
compared with the escape energy.

\begin{figure}[t]
\begin{center}
 \includegraphics[height=6.5cm]{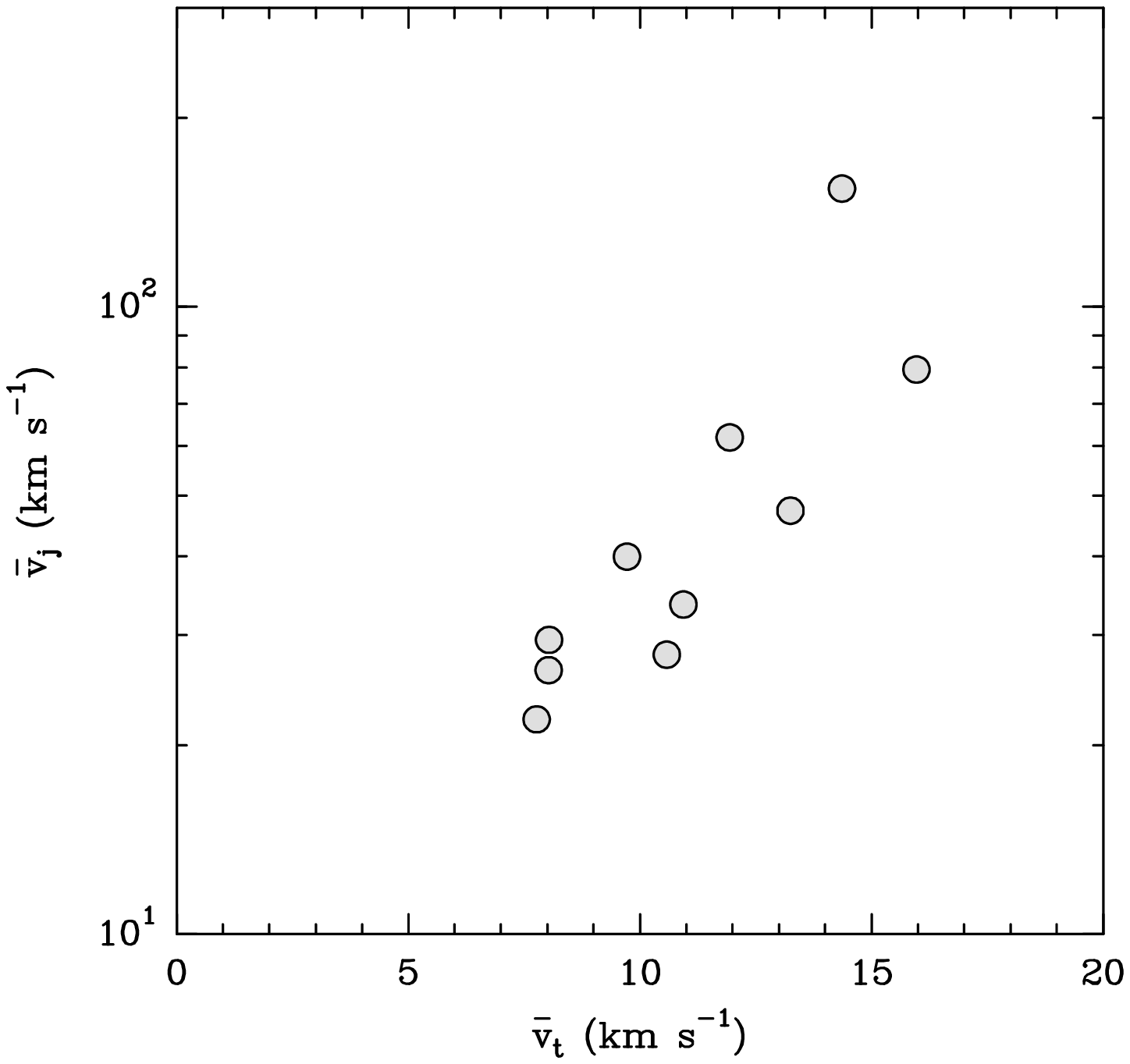}
\hspace{0.2cm} \includegraphics[height=6.5cm]{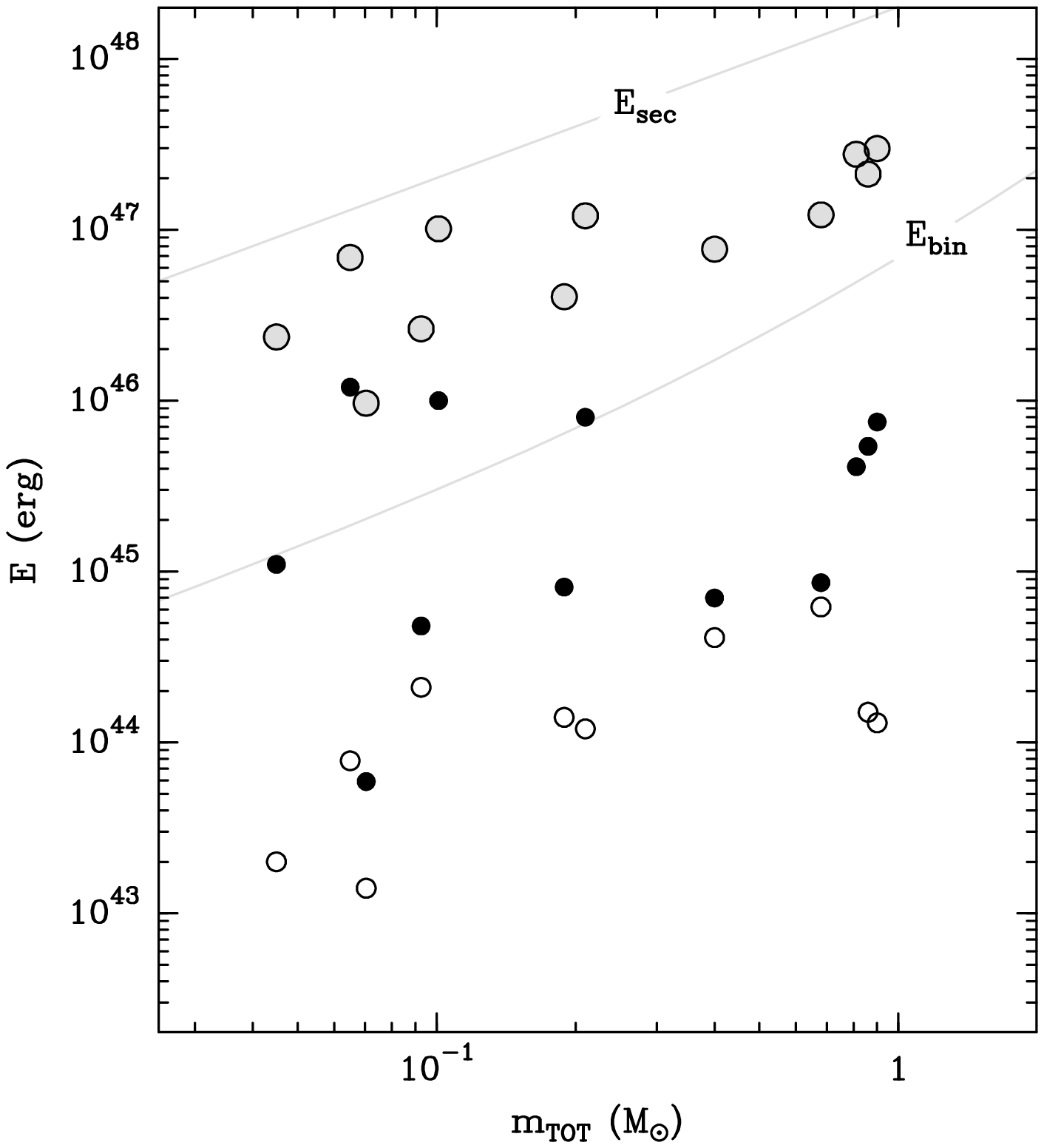}
 \caption{\emph{Left}: Mean velocities of the jets vs.\ mean velocities of
the tori. \emph{Right}: Kinetic energies of the jets (filled circles)
and tori (filled circles), and estimated input energies (gray-filled
circles). The curves labeled $E_\mathrm{sec}$ and $E_\mathrm{bin}$
are reference energies described in the text.}
   \label{fig1}
\end{center}
\end{figure}

{\underline{\it Input energies}}.
The currently observed kinetic energy in a pre-PN is much less than
the initially injected kinetic energy, for two reasons. First the gas
has done work against the gravitational field of the central star; and
second, the interactions of the initial jets in forming the outflows
is highly inelastic, and much of the energy is lost in the form of
radiation. We focus on the second effect which is larger.

Because the momentum of the outflows is conserved and is directly
observed, we can use the momentum to estimate the initial energy
($E_\mathrm{o}$) if we know the initial velocity. Fig.~1 shows our
estimates of the total energy input as gray-filled circles, based on a
nominal initial velocity 1350~km s$^{-1}$ (three times the Keplerian
velocity of lower mass main sequence stars). The actual initial
velocity is likely to be within a factor of a few of this number;
larger for core accretion, and at least several 100~km~s$^{-1}$ based
directly on the observed jet spectra. It can be seen that
$E_\mathrm{o}>E_\mathrm{bin}$ and $E_\mathrm{o}<E_\mathrm{sec}$.

\section{Observations vs.\ theory}
The dynamical properties of the ensemble of pre-PNe described here
place useful constraints on possible ejection mechanisms. We consider
the three most widely discussed scenarios, all of which involve a companion.

{\underline{\it Accretion onto the AGB core}}. The first is based on
accretion of part or all of a companion onto the dense core of the AGB
star or its remnant, either during a common envelope phase or later
when the envelope has been removed (e.g., \cite[Soker \& Livio
1994]{sok94}).  The energy generated by such a process depends mainly
on the accreted mass, which in turn depends on the
detailed circumstances of the accretion. It seems unlikely that such a
model could inject energy for the sample of pre-PNe at a level that
depends on the envelope mass, as found here, and there are no
predictions of such an an effect.  This scenario must remain doubtful,
and a variation with long time-scales has already been ruled out by
timing arguments (\cite[Huggins 2007]{hug07}).

{\underline{\it Infall}}. The infall of a companion is consistent with
the energetics of the tori, but standard common envelope
calculations do not produce high velocity jets. Other mechanisms have
been suggested to tap the infall energy (e.g., \cite[Nordhaus \&
Blackman 2006]{nor06}), but there are no detailed proposals that
produce directed flows with high efficiency and energies in excess
of the binding energy. Thus jet scenarios driven by infall remain
unproven.

{\underline{\it Accretion onto a companion}}.  The observed
energetics of the jets and tori are consistent with the third
scenario, the accretion onto a companion star (e.g.,
\cite[Morris 1987]{mor87}, \cite[Soker \& Rappaport 2000]{sok00}).  In
addition, consideration of the observed momentum in the flows and the
efficiency of momentum production based empirically on YSO jets or
disk-jet theory demonstrates that the accretion of a few tenths of the
ejected envelope can give rise to the observed momentum.  This
particular scenario is strengthened by the fact that the few known AGB
stars with tori and jets follow the same momentum-mass relation.  One
of these AGB stars (V~Hya) is a known, detached binary with
spectroscopic evidence for accretion onto a companion, and one of the
sample pre-PNe (HD~101584) is a binary, and has evidently passed
through a mild common envelope phase. Some of the other sample pre-PNe have
observational limits on possible close companions (\cite[Hrivnak et
al.\ 2011]{hri11}). It will be of considerable interest to see if
these observations can be refined, to strengthen or challenge our
conclusions.

\begin{acknowledgments}
This work is supported in part by NSF grant AST 08-06910.
\end{acknowledgments}

\end{document}